\documentclass[preprint]{aastex63}

\usepackage{amsmath}
\usepackage{color}
\definecolor{DarkGreen}{rgb}{0.0,0.45,0.0}  
\newcommand{\B}[1]{{\color{blue}{#1}}}      
\newcommand{\R}[1]{{\color{red}{#1}}}     
\newcommand{\G}[1]{{\color{DarkGreen}{#1}}}
\newcommand{\M}[1]{{\color{magenta}{#1}}}

\submitjournal{ApJ}

\begin{document}

\title{Electric Currents through J-shaped and Non-J-shaped Flare Ribbons}

\author{Yuwei He}
\affiliation{CAS Key Laboratory of Geospace Environment, Department of Geophysics and Planetary Sciences, University of Science and Technology of China, Hefei, Anhui 230026, China}

\author[0000-0003-4618-4979]{Rui Liu}
\affiliation{CAS Key Laboratory of Geospace Environment, Department of Geophysics and Planetary Sciences, University of Science and Technology of China, Hefei, Anhui 230026, China}
\affiliation{CAS Center for Excellence in Comparative Planetology, Hefei, Anhui 230026, China}
\affiliation{Mengcheng National Geophysical Observatory，University of Science and Technology of China, Mengcheng, Anhui 233500, China}

\author[0000-0001-6804-848X]{Lijuan Liu}
\affiliation{School of Atmospheric Sciences, Sun Yat-sen University, Zhuhai, Guangdong 519082, China }
\affiliation{CAS Center for Excellence in Comparative Planetology, Hefei, Anhui 230026, China}

\author[0000-0003-3060-0480]{Jun Chen}
\affiliation{CAS Key Laboratory of Geospace Environment, Department of Geophysics and Planetary Sciences, University of Science and Technology of China, Hefei, Anhui 230026, China}

\author[0000-0002-9865-5245]{Wensi Wang}
\affiliation{CAS Key Laboratory of Geospace Environment, Department of Geophysics and Planetary Sciences, University of Science and Technology of China, Hefei, Anhui 230026, China}

\author[0000-0002-8887-3919]{Yuming Wang}
\affiliation{CAS Key Laboratory of Geospace Environment, Department of Geophysics and Planetary Sciences, University of Science and Technology of China, Hefei, Anhui 230026, China}
\affiliation{CAS Center for Excellence in Comparative Planetology, Hefei, Anhui 230026, China}

\correspondingauthor{Rui Liu}
\email{rliu@ustc.edu.cn}
	
\begin{abstract}
Recently solar flares exhibiting a double J-shaped ribbons in the lower solar atmosphere have been paid increasing attention in the context of extending the two-dimensional standard flare model to three dimensions, as motivated by the spatial correlation between photospheric current channels and flare ribbons. Here we study the electric currents through the photospheric area swept by flare ribbons (termed synthesized ribbon area or SRA), with a sample of 71 two-ribbon flares, of which 36 are J-shaped. Electric currents flowing through one ribbon are highly correlated with those through the other, therefore belonging to the same current system. The non-neutrality factor of this current system is independent of the flare magnitude, implying that both direct and return currents participate in flares. J-shaped flares are distinct from non-J-shaped flares in the following aspects: 1) Electric current densities within J-shaped SRA are significantly smaller than those within non-J-shaped SRA, but J-shaped SRA and its associated magnetic flux are also significantly larger. 2) Electric currents through SRA are positively correlated with the flare magnitude, but J-shaped flares show stronger correlation than non-J-shaped flares. 3) The majority (75\%) of J-shaped flares are eruptive, while the majority (86\%) of non-J-shaped flares are confined; accordingly, hosting active regions of J-shaped flares are more likely to be sigmoidal than non-J-shaped flares. Thus, J-shaped flares constitute a distinct subset of two-ribbon flares, probably the representative of eruptive ones. Further, we found that combining SRA and its associated magnetic flux has the potential to differentiate eruptive from confined flares. 

\end{abstract}
	
\section{Introduction}
	
Solar flares are among the most energetic phenomena in our solar system. The flare emission spans the whole range of electromagnetic spectrum, often associated with particle acceleration and coronal mass ejections. The physical mechanism of solar flares is not only important for forecasting the space weather at Earth, but for understanding similar physical processes in stellar flares as well as on planetary magnetospheres \citep[e.g.,][]{Zhong2020} and active galactic nuclei \citep[e.g.,][]{Barret&Cappi2019}. Although the complexity and diversity of flare phenomena makes it impossible to build a `universal' flare model that is capable of explaining all observational aspects in all events, the standard or CSHKP flare model \citep{KoppRA1976, Sturrock1966, Carmichael1964, Hirayama1974} is successful in explaining the major characteristics of two-ribbon eruptive flares, and has formed the basis for our understanding of solar flares for decades \citep{Priest&Forbes2002}. 

Solar eruptive phenomena draw energy from coronal magnetic fields \citep{Forbes2000}. The magnetic field $\mathbf{B}$ can be always decomposed into a current-free, potential component $\mathbf{B}_p$ and a current-carrying, non-potential component $\mathbf{B}_c$, so that the magnetic energy $E_m$ in a volume $V$ can be written as \citep{Sakurai1981}
\begin{equation}
E_m=\int_V \frac{B^2}{8\pi}\,dV=\frac{1}{8\pi}\int_VB_p^2\,dV+\frac{1}{2c}\int_V \mathbf{A}_c\cdot\mathbf{J}\,dV,
\end{equation} where $\mathbf{B}_c=\nabla\times\mathbf{A}_c$ and $\mathbf{J}=\frac{c}{4\pi}\nabla\times\mathbf{B}_c$, as $\nabla\times\mathbf{B}_p=0$. The first term is the energy of the potential field, which is produced by current sources located within the interior of the Sun. The energy powering solar eruptions can only be contained in the second term that is associated with electric currents in the solar corona. The most rapid release of this `free' magnetic energy is manifested as flares. Because of the huge magnetic Reynolds number in the corona, electric currents must be concentrated into small regions such as narrow current sheets, across which magnetic connectivity changes rapidly, generally known as quasi-separatrix layers \citep[QSLs;][]{Demoulin2006}. QSLs are preferential sites for magnetic reconnection, which drives the processes of plasma heating and particle acceleration in flares. Due to the requirement of the electric current continuity, concentrations of electric currents in the photosphere are believed to be the imprints of coronal current sheets \citep{Fleishman&Pevtsov2018}. During the flare, the release of free magnetic energy stored in the corona can be attributed to a geometric reconfiguration of the current paths, with the magnitudes of currents at the footpoints of the current system being fixed to a large extent \citep{Melrose2017}. During some flares, however, horizontal electric currents tend to concentrate at lower altitudes around the polarity inversion line (PIL) than before the eruption \citep[e.g.,][]{Sun2012, Liu2012,Liu2014}, while vertical electric currents tend to increase in localized ribbons \citep[e.g.,][]{Janvier2014,Janvier2016,Sharykin2020}.

From the observational perspective, only the vertical component of the electric current density $J_z$ can be derived from photospheric vector magnetograms, which are normally limited to a single height. With the measurements of photospheric transverse magnetic fields becoming more and more reliable, it has been well known that there exists a close spatial relationship between the vertical electric currents at the photosphere and the deposition sites of flare energy at the chromosphere as represented by bright H$\alpha$ kernels \citep[e.g.,][]{Moreton1968,Yuanzhang1987,Romanov1990,Leka1993,deLaBeaujardiere1993,Sharykin2014}, or UV/EUV emission \citep[e.g.,][]{Janvier2014,Janvier2016,Sharykin2020}, or hard X-ray emission \citep[e.g.,][]{deLaBeaujardiere1993, Canfield1992, lijing1997,Musset2015,Sharykin2020}. However, these flare kernels are often not exactly co-spatial with the locations of strongest electric current densities, but adjacent to the current channels  \citep[e.g.,][]{Romanov1990,Leka1993,deLaBeaujardiere1993,Canfield1992,lijing1997}, which might result from magnetic reconnection in a quadrupolar magnetic field \citep{Aschwanden1999}. Thus, the distribution of pre-flare electric current densities in the photosphere provides important clues to the coronal currents accessible to flares.

More recently, aided by nonlinear force-free field or MHD modeling of the coronal magnetic field, it has been demonstrated that H$\alpha$ and UV flare ribbons often coincide with the footprints of QSLs \citep[e.g.,][]{Liu2014, Liu2016SR, Janvier2014, Janvier2016, Liu2018, Su2018, Jiang2018}. In particular, the footprints of the QSLs wrapping around a magnetic flux rope, the core structure of coronal mass ejections \citep{Vourlidas2013,Georgoulis2019}, correspond to a pair of J-shaped ribbons of high electric current densities; the hooks of J-shaped ribbons outline the edge of the rope legs \citep{Janvier2014,Janvier2016,Wang2017}. Motivated by these observational and modeling results, it has been proposed that the two-dimensional standard model can be extended to three dimensions to address the shape, location, and dynamics of flares with a double J-shaped ribbons \citep{Aulanier2013, Aulanier2012, Janvier2013, Janvier2015}. However, not all two-ribbon flares are J-shaped; in fact, many typical two-ribbon flares \citep[e.g., Figure 1 in][]{Qiu2010} do not exhibit hooks at the ends of flare ribbons. Hence, it is obscure whether a canonical J-shaped flare, which breaks the translational symmetry along ribbons but introduce a 2-fold rotational symmetry (i.e., the ribbon morphology does not change by a rotation of $180^\circ$), can represent classic two-ribbon flares in the general 3D context.  

In this paper, we tackle on this question by investigating the distribution of pre-eruption photospheric electric currents associated with flare ribbons. Since the two ribbons often move away from each other during the impulsive phase of flares, we consider the photospheric electric currents through the total area swept by flare ribbons, termed synthesized ribbon area (SRA) hereafter, instead of the area covered by flare ribbons at any time instant. In the rest of the paper, we present the methods in \S\ref{sec:method} and the statistical analysis in \S\ref{sec:statistics}, and make concluding remarks in \S\ref{sec:conclusion}.

\section{Data and Methods} \label{sec:method}
\subsection{Selection of Events}
\begin{figure}[htbp]
	\plotone{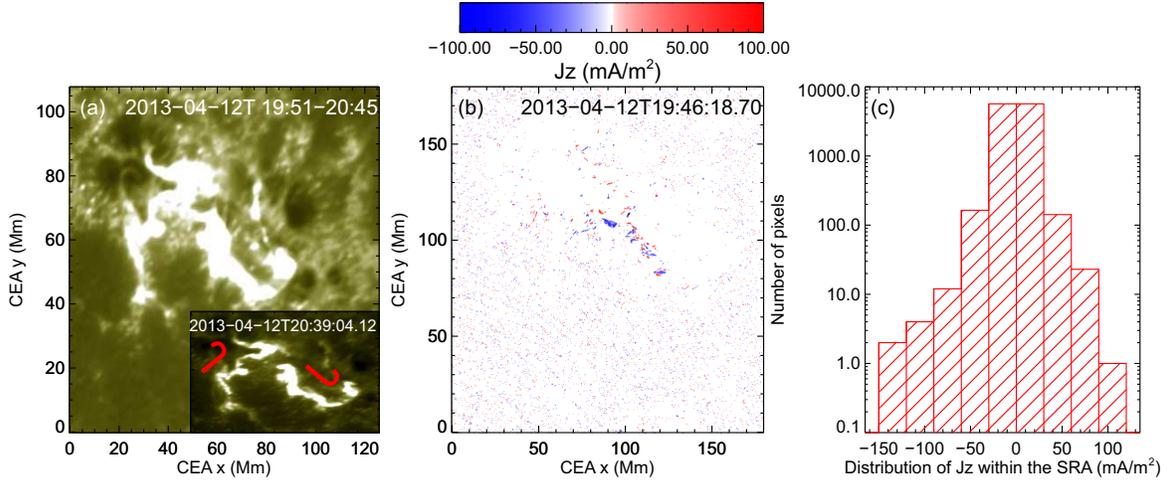}
	\caption{An exemplary J-shaped flare observed on 2013 April 12. a) AIA 1600~{\AA} emission averaged over from the flare onset to end time; the image is projected to CEA coordinates, and the inset shows flare ribbons at a time instant. The J shapes are illustrated by two reverse J's. b) Distribution of pre-flare $J_z$ in the same field of view as Panel a) at 19:46~UT, in which pixels with $|J_z|$ below 22~mA~m$^{-2}$ are assigned zero values, hence shown in white. c) Histograms of $J_z$ within the synthesized ribbon area, including $|J_z|$ values below 22~mA~m$^{-2}$. \label{ffg:J-ribbon} }
\end{figure}

\begin{figure}[htp!]
	\plotone{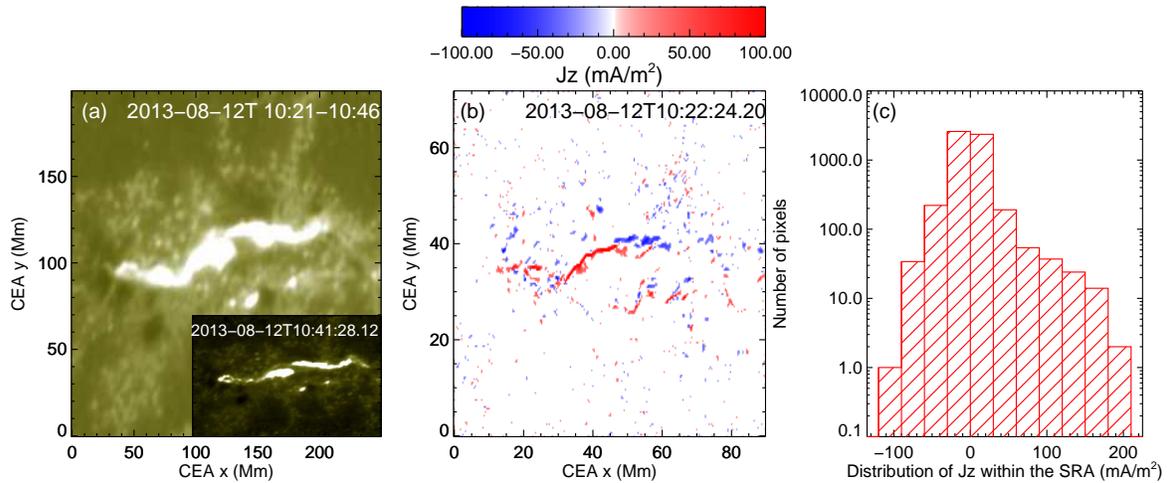}
	\caption{An exemplary non-J-shaped flare observed 2013 August 12. a) AIA 1600~{\AA} emission averaged over from the flare onset to end time; the image is projected to CEA coordinates, and the inset shows flare ribbons at a time instant. b) Distribution of pre-flare $J_z$ in the same field of view as Panel a) at 10:22~UT, in which pixels with $|J_z|$ below 22~mA~m$^{-2}$ are assigned zero values, hence shown in white. c) Histograms of $J_z$ within the synthesized ribbon area, including $|J_z|$ values below 22~mA~m$^{-2}$. \label{ffg:non-J-ribbon}}
\end{figure}

We selected two-ribbon flares of GOES-class M1.0 and above from the database provided by \citet{Kazachenko2017}. The database covers all flare ribbon events between 2010 April and 2016 April recorded by the Solar Dynamics Observatory \citep[SDO;][]{Pesnell2012}, corresponding to GOES-class C1.0 and above within 45$^\circ$ from the central meridian. By visually examining flare ribbons observed by the ultraviolet 1600~{\AA} passband of the Atmospheric Imaging Assembly \citep[AIA;][]{Lemen2012} onboard SDO, we categorized the selected events into J-shaped flares (Figure~\ref{ffg:J-ribbon}a), if at least one of the two ribbons exhibits a hooked shape at the ribbon end, and non-J-shaped flares (Figure~\ref{ffg:non-J-ribbon}a), if neither ribbon has a pronounced hook. It is interesting that J-shaped flares do not necessarily occur in a sigmoidal active region (indicated by an asterisk in Tables~\ref{tab:Non-J-shaped flares} and \ref{tab:J-shaped flares}), which exhibits an overall S shape with two opposing bundles of coronal loops; but non-J-shaped flares can sometimes be hosted by a sigmoidal active region. In addition, we excluded two types of events: 1) those with material ejection visible near flare ribbons in AIA 1600~{\AA} images; and 2) those with complex or remote ribbons beyond two major ribbons in AIA 1600~{\AA}. Both scenarios could compromise a clear identification of the two major ribbons. In total, we have 36 J-shaped and 35 non-J-shaped flares. 	

\subsection{Calculation of Electric Current Densities}
We calculated the distribution of current density $J_z$ in active regions of interest by Amep\'{e}re's law, 
\begin{equation}
\mu_0J_z= \frac{\partial B_y}{\partial x} - \frac{\partial B_x}{\partial y}.
\end{equation}
$B_{x,y}$ is obtained by the Helioseismic and Magnetic Imager \citep[HMI;][]{Scherrer2012} onboard SDO. For each flare, we worked with a vector magnetogram taken immediately before the flare onset. These vector magnetograms are disambiguated and de-projected to the heliographic coordinates with a Lambert (cylindrical equal area; CEA) projection method, also known as the Space-Weather HMI Active Region Patches (SHARP) data. Computation of electric currents based on the above equation is subject to several uncertainties that are difficult to quantify, e.g., the 180$^\circ$ ambiguity in the horizontal field direction, and these uncertainties can be further amplified by partial derivatives. Considering the uncertainty in measuring the transverse component of photospheric magnetic field \citep[$\sim\,$100 G;][]{Hoeksema2014} and SHARP's pixel size (0.36 Mm), we estimated that the uncertainty of $J_z$ is about 22~mA~m$^{-2}$. As a comparison, the average and standard deviation of $|J_z|$ in a box area in the quiet region is listed in the column `noise' in Tables~\ref{tab:Non-J-shaped flares} and \ref{tab:J-shaped flares}, and the box typically occupies the lower left corner (1/12 in length and 1/8 in width) of the SHARP map. One can see that in all cases the value of 22~mA~m$^{-2}$, which significantly exceeds the average $|J_z|$ in the quiet region ($\sim\,$10~mA~m$^{-2}$), is a reliable representative of the noise level. Hence we considered only $J_z$ above 22~mA~m$^{-2}$ in the statistical analysis below (\S\ref{sec:statistics}). We have also carried out the same analysis without considering this threshold value, but got similar results and reached the same conclusions.

\subsection{Identification of Flare Ribbons}
We used AIA 1600~{\AA} images to identify flare ribbons. The AIA 1600~{\AA} passbands provides full-disk images of the lower atmosphere at a temporal cadence of 24~s and a spatial resolution of $1''.2$. To match the map of electric current density with that of flare ribbons, we remapped 1600~{\AA} images from the CCD coordinates to the CEA coordinates of the corresponding SHARP data. We detected flare ribbons by setting a threshold of about 3$\sim$5 times the average brightness of an AIA 1600~{\AA} image immediately before the onset of each flare under investigation. We fine-tuned the exact threshold value case by case so that flare ribbons are visually captured as accurately as possible while as many as bright plages are excluded. This threshold value is varied by 15\% to be taken as the upper and lower threshold. We flagged the pixels with brightness above the upper/lower threshold in each 1600~{\AA} image, and collected all the flagged pixels during the interval from the GOES flare start to end time to construct `synthesized' flare ribbons. Meanwhile we removed isolated pixels flagged outside the main ribbon area, and then constructed a binary mask for the synthesized flare ribbons by setting the flagged pixels to be unity and the rest to be zero. Applying the upper and lower threshold yields two slightly different ribbon masks and different ribbon areas. The average area is taken as the synthesized ribbon area (SRA). In the following, when calculating an electric-current parameter, we also apply the two ribbon masks that are associated with the upper and lower threshold, respectively, and then take the average to be the final result and half of the range to be the uncertainty.

\subsection{Calculation of Electric Current Parameters}
\begin{figure}[htp!]
	\plotone{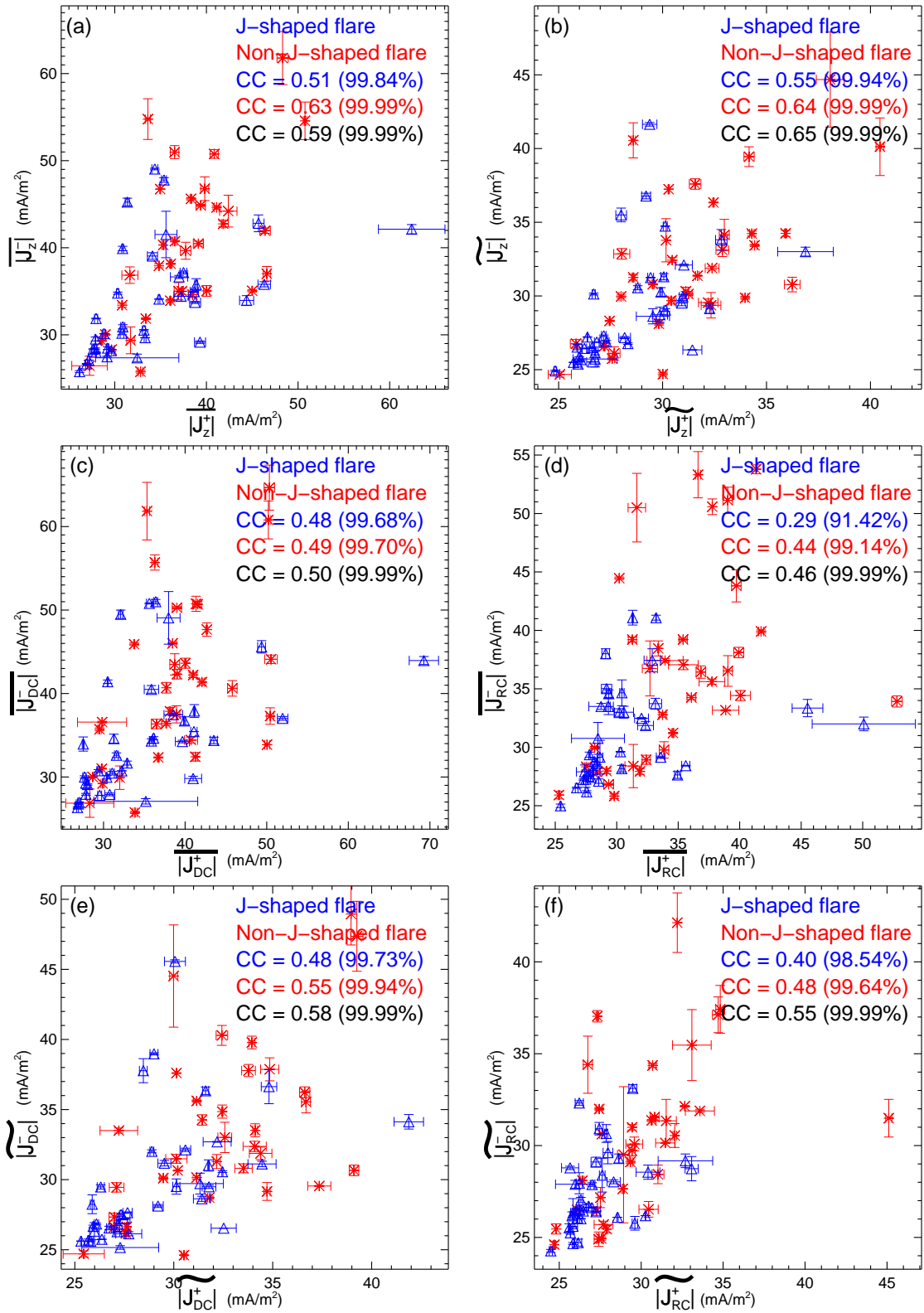}
	\caption{Electric current densities within the synthesized ribbon of positive polarity vs. negative polarity. Red asterisks and blue triangles represent Non-J-shaped and J-shaped flares, respectively. Annotated in each panel are the Pearson correlation coefficient (CC) and corresponding confidence level (in the brackets) for each category (color coded) and the whole sample (black). \label{fig:ribbon_vs_ribbon_J}}
\end{figure}

\begin{figure}[htp!]
	\plotone{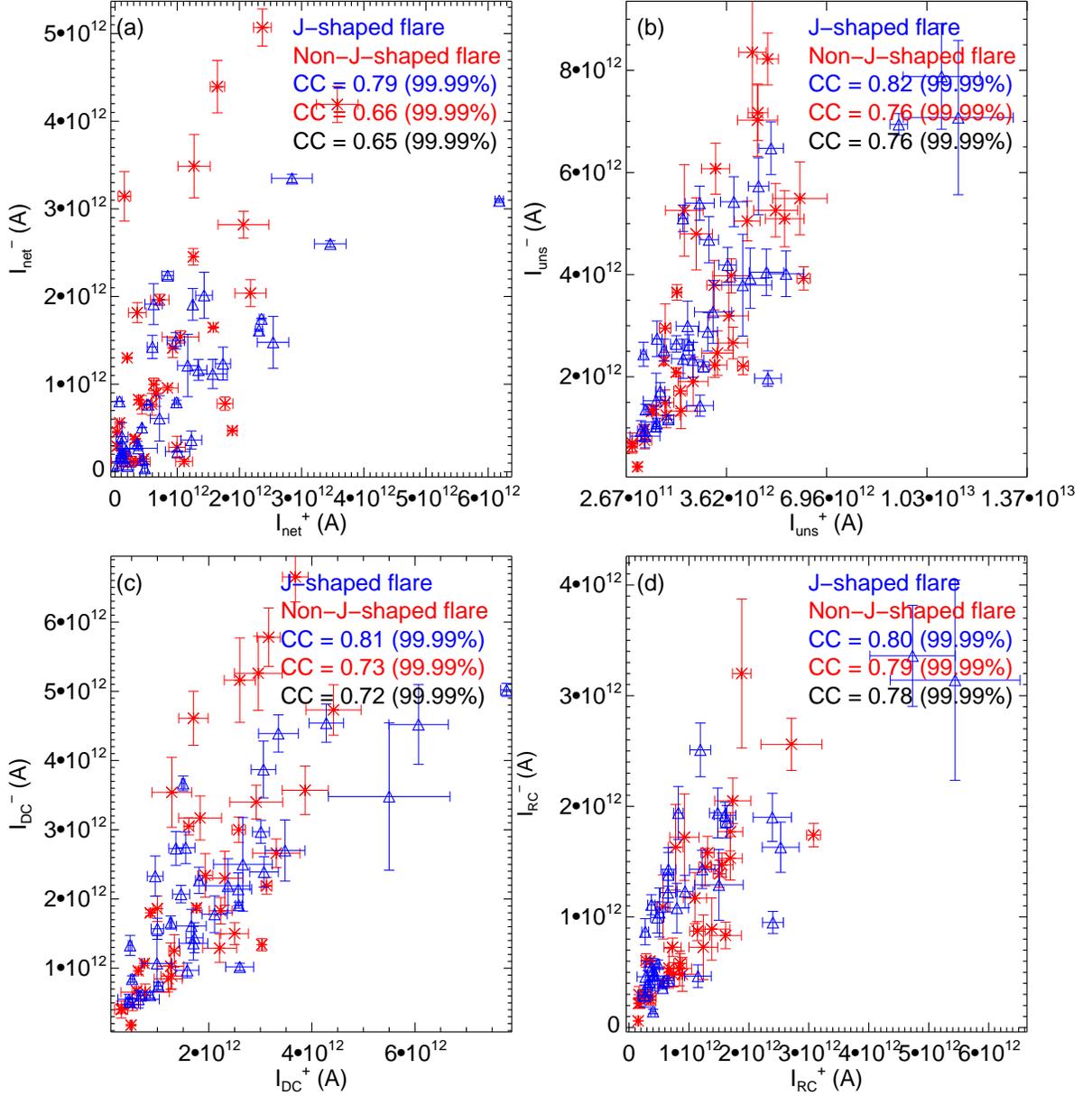}
	\caption{Electric currents through the synthesized ribbon of positive polarity vs. negative polarity. Red asterisks and blue triangles represent Non-J-shaped and J-shaped flares, respectively. Annotated in each panel are the Pearson correlation coefficient (CC) and corresponding confidence level (in the brackets) for each category (color coded) and the whole sample (black).} \label{fig:ribbon_vs_ribbon_I}
\end{figure}

Multiplying the map of pre-flare $J_z$ by the synthesized flare-ribbon mask, we obtained the distribution of $J_z$ within SRA (Figures~\ref{ffg:J-ribbon}c and \ref{ffg:non-J-ribbon}c) as well as the mean ($\overline{|J_z|}$) and median ($\widetilde{|J_z|}$) of the absolute values of $J_z$. The direct current density $J_\mathrm{DC}$ is selected from $J_z$ whose sign is consistent with the dominant current through the ribbon, as opposed to the return current density $J_\mathrm{RC}$. Note that for simplicity, here the signs of direct and return currents are not defined according to the sign of magnetic helicity of the host active region as in MHD simulations \citep{Schmieder&Aulanier2018}. In observation, it is more difficult to determine the sign of helicity than that of dominant current. Summing up $J_\mathrm{DC}$ and $J_\mathrm{RC}$ on a flare ribbon, we obtained the direct current $I_\mathrm{DC}$ and return current $I_\mathrm{RC}$, respectively. The net current $I_\mathrm{net}$ is the algebraic sum of $J_z$ over SRA, as opposed to the unsigned current $I_\mathrm{uns}$, which is the sum of $|J_z|$. The above quantities are listed in Tables~\ref{tab:Non-J-shaped flares} and \ref{tab:J-shaped flares}. We adopted the superscript `+' or `-' to indicate that a parameter is given for a ribbon located on the positive or negative polarity side of the PIL. When the superscript is dropped, electric current densities are sampled through both ribbons.

The above defined parameters of electric current densities (Fig.~\ref{fig:ribbon_vs_ribbon_J}) and currents (Fig.~\ref{fig:ribbon_vs_ribbon_I}) flowing through the SRA of positive polarity are highly correlated with that of negative polarity. Hence it is reasonable to assume that they belong to the same current system, i.e., the electric currents flow out of one ribbon and into the other. Thus, following \citet{Georgoulis2012} we defined a non-neutrality factor $I_\mathrm{nn}$, 
\begin{equation} \label{eq:Inn}
I_\mathrm{nn}=\frac{1}{2}\left(\frac{|I_\mathrm{net}^+|}{|I_\mathrm{DC}^+|}+\frac{|I_\mathrm{net}^-|}{|I_\mathrm{DC}^-|}\right),
\end{equation}
so that $I_\mathrm{nn}=0$ if electric currents through the SRA are balanced separately on either side of the PIL, i.e., neutralized, and that $I_\mathrm{nn}=1$ if electric currents through the SRA are unneutralized. We noticed that \cite{LiuY2017} used $|I_\mathrm{DC}/I_\mathrm{RC}|$ to measure the degree of current neutralization in active regions, but this parameter is less tractable than $I_\mathrm{nn}$ when $I_\mathrm{RC}$ is small.    

\section{Statistics}\label{sec:statistics}
\begin{figure}[htp!]
	\plotone{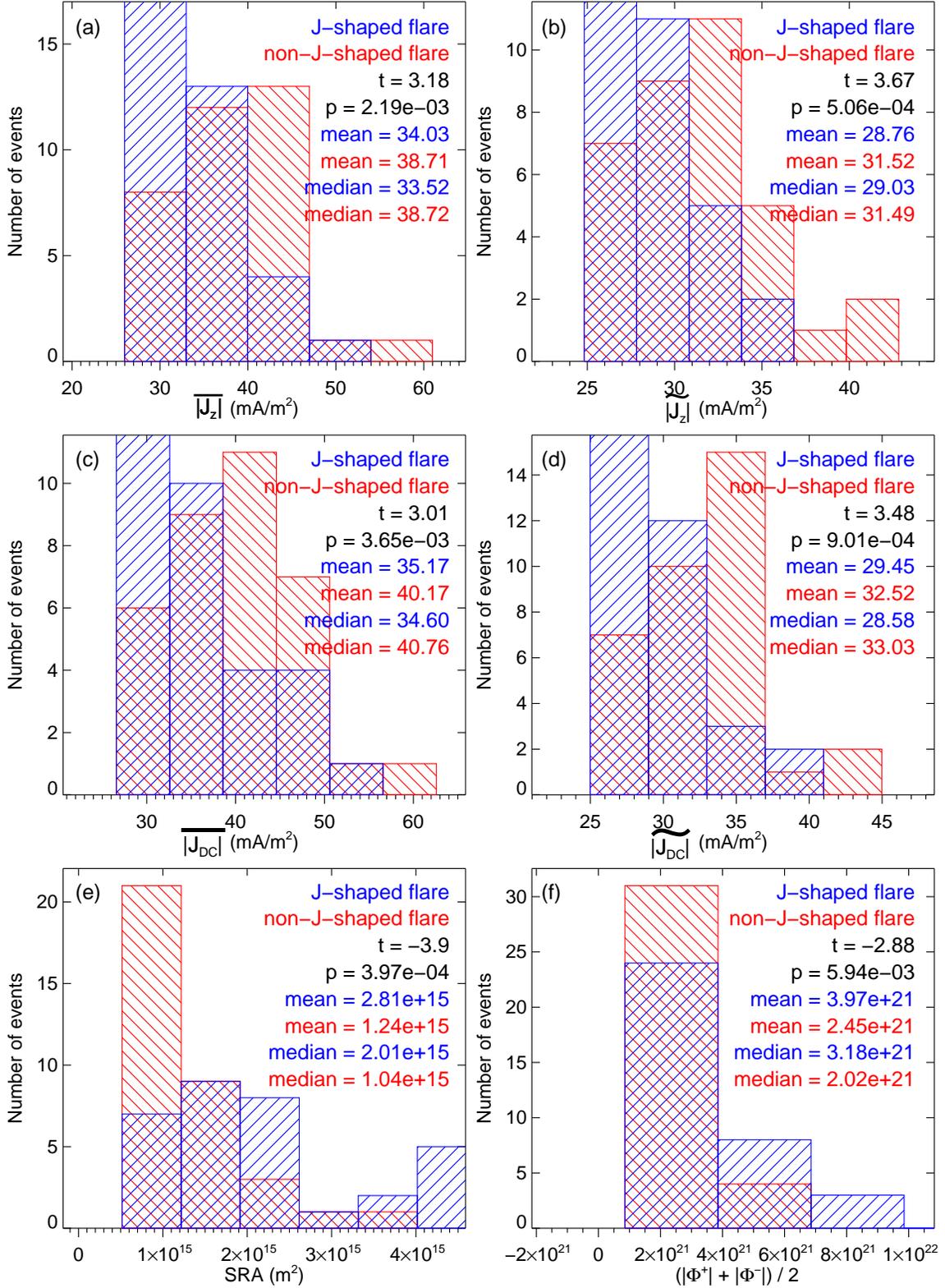}
	\caption{Histograms of parameters that display statistically significant differences between J-shaped and non-J-shaped flares. (a \& b) The mean and median of $|J_z|$ within the SRA, respectively; (c \& d) the mean and median of $|J_\mathrm{DC}|$ within the SRA, respectively; (e) the distribution of SRA; (f) the distribution of unsigned magnetic flux through SRA, averaged over two ribbons. J-shaped (non-J-shaped) flares are indicated by blue (red) colors. Welch's statistic $t$ and the corresponding $p$-value are annotated in each panel. \label{fig:student_test}}
\end{figure}

\begin{figure}[htp!]			
	\plotone{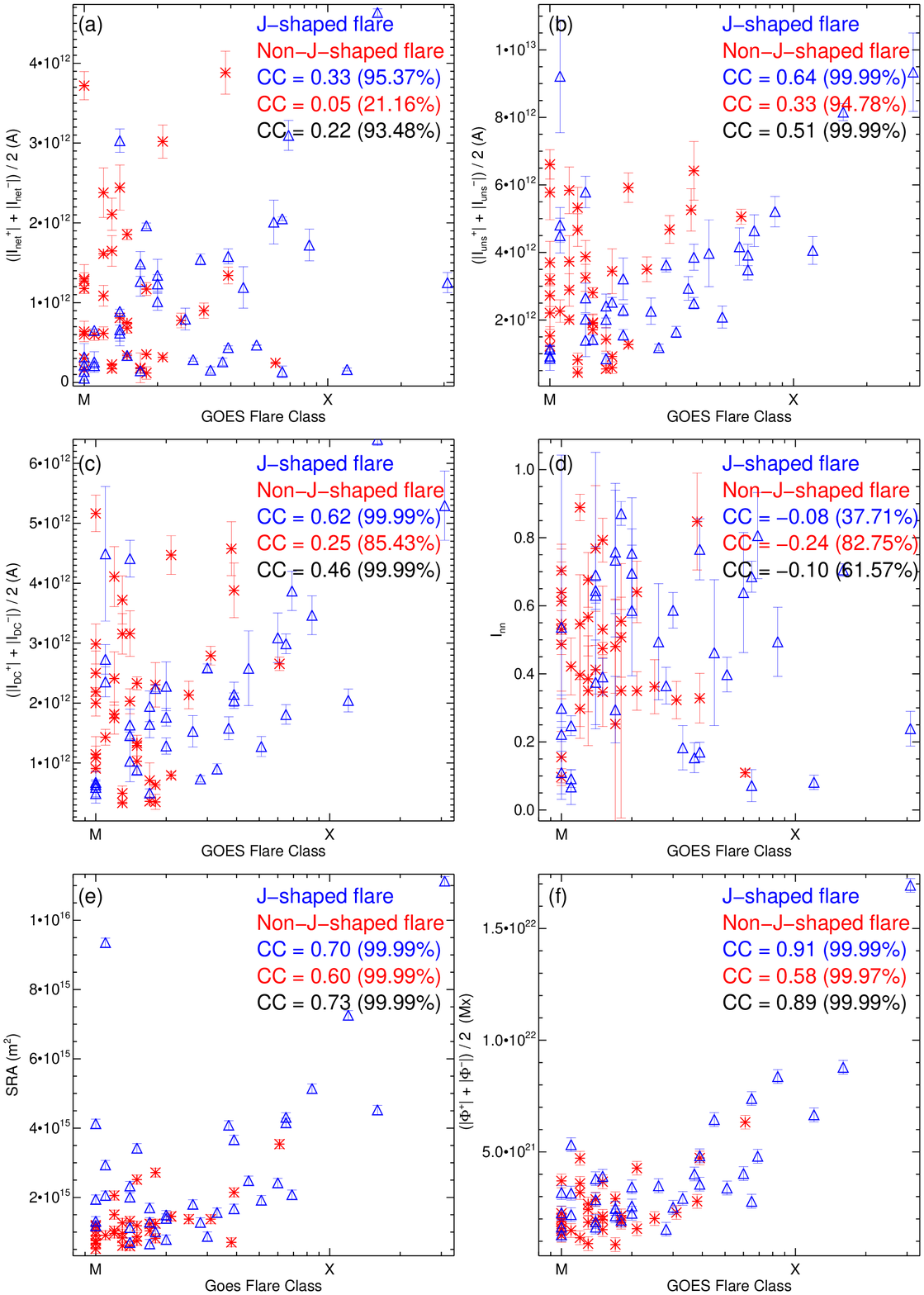}
	\caption{Electric currents and magnetic flux through synthesized ribbon area in relation to GOES flare class. 
		Plotted are (a) net current averaged over two ribbons, (b) unsigned current averaged over two ribbons, (c) direct current averaged over two ribbons, (d) non-neutrality factor as defined by Eq.~\ref{eq:Inn}, (e) synthesized ribbon area, and (f) magnetic flux through SRA averaged over two ribbons. Red asterisks and blue triangles represent Non-J-shaped and J-shaped flares, respectively. Annotated in each panel are the Pearson correlation coefficient (CC) and corresponding confidence level (in the brackets) for each category (color coded) and the whole sample (black).} \label{fig:I_vs_goes}. 
\end{figure}	

To evaluate whether J-shaped flares are distinct from non-J-shaped flares in terms of electric currents through SRA, we performed Welch's $t$-test to determine whether two samples of different sizes and variances belong to the same population, by comparing their sample means. The test defines a statistic $t$ as follows, 
\begin{equation}
t=\frac{\overline{X}_1-\overline{X}_2}{\sqrt{\frac{s_1^2}{N_1}+\frac{s_2^2}{N_2}}}, 
\end{equation} where $\overline{X}_1$, $s_1^2$, and $N_1$ are the 1st sample mean, sample variance, and sample size, respectively. 
The null hypothesis that two populations have equal means can be rejected if the resultant $p$-value is small enough. As a rule of thumb, one may deem that the two populations have significantly different means if $p\le 0.01$, as there is less than a 1\% probability that the null hypothesis is correct.  

We found that the distributions of most current-density parameters are different for the two flare categories, which is manifested by the statistical significance in Welch's $t$-test. Specifically, both median and mean $|J_z|$ ($\widetilde{|J_z|}$ and $\overline{|J_z|}$, respectively) through J-shaped ribbons are smaller than those through non-J-shaped ribbons (Figure~\ref{fig:student_test}(a \& b)). A similar result is found for the mean and median current densities for DC (Figure~\ref{fig:student_test}(c \& d)) and RC (not shown). But we found no statistically significant differences between J-shaped and non-J-shaped flares in electric currents through SRA, no matter if it is the direct, return, unsigned, or net current, or the dominance of direct current as measured by $I_\mathrm{DC}^+/I_\mathrm{RC}^+$ and $I_\mathrm{DC}^-/I_\mathrm{RC}^-$, or the current imbalance as measured by  $I_\mathrm{DC}^+/I_\mathrm{DC}^-$ and $I_\mathrm{RC}^+/I_\mathrm{RC}^-$. Note all the ratios are taken in their absolute values. However, J-shaped flares have significantly larger SRA (Figure~\ref{fig:student_test}e) as well as larger unsigned magnetic flux through it $(|\Phi^+| + |\Phi^-|)/2$ (averaged over two ribbons; Figure~\ref{fig:student_test}f) than non-J-shaped flares. 

We further examined the correlations between the flare magnitude as measured by the GOES class and the above parameters. None of the current density parameters, $|J_z|$, $|J_\mathrm{RC}|$, or $|J_\mathrm{DC}|$ within SRA, shows significant correlation with the flare magnitude. Averaged over two ribbons, the mean net current $(|I_\mathrm{net}^+|+|I_\mathrm{net}^-|)/2$, mean unsigned current $(|I_\mathrm{uns}^+|+|I_\mathrm{uns}^-|)/2$, and mean direct current $(|I_\mathrm{DC}^+|+|I_\mathrm{DC}^-|)/2$ are all positively correlated with the flare magnitude; but the correlation coefficient for J-shaped flares is consistently larger than non-J-shaped flare (Figure~\ref{fig:I_vs_goes}(a--c)). The flare magnitude is also positively correlated with both SRA (Figure~\ref{fig:I_vs_goes}e) and magnetic flux through SRA $(|\Phi^+| + |\Phi^-|)/2$ (Figure~\ref{fig:I_vs_goes}f), which is consistent with \citet{Kazachenko2017} and previous studies. On the other hand, it is surprising that the non-neutrality factor $I_\mathrm{nn}$ is almost independent of the flare magnitude (Figure~\ref{fig:I_vs_goes}d). 

\section{Discussion and Conclusion} \label{sec:conclusion}
\begin{figure}[htp!]			
	\plotone{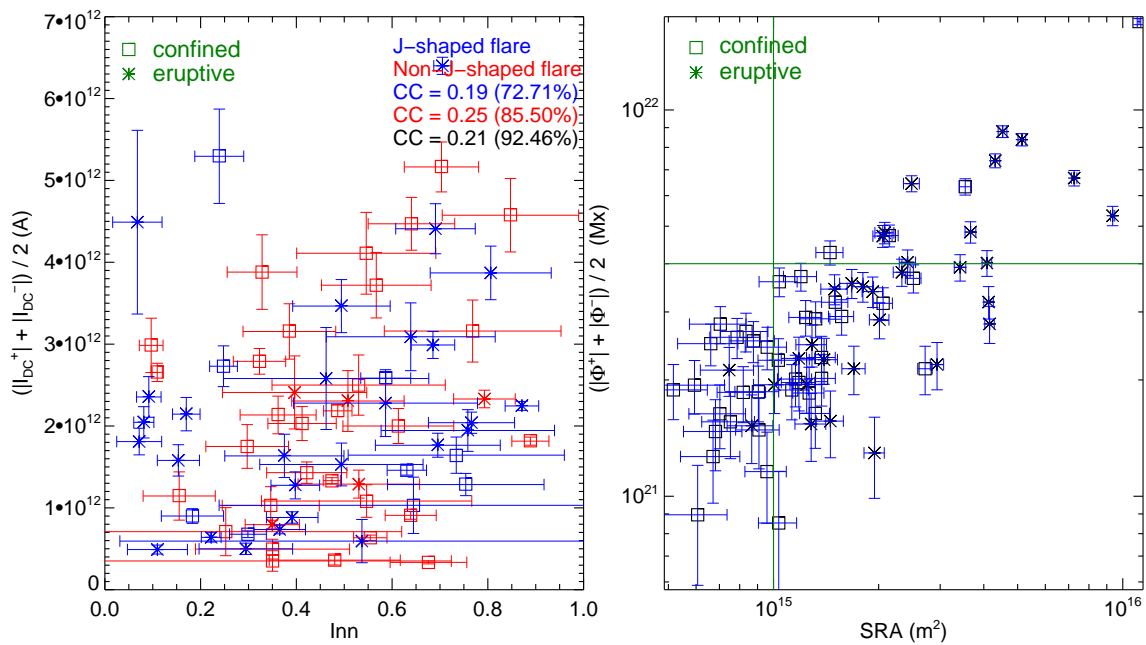}
	\caption{Eruptive and confined flares in different parameter spaces. Eruptive flare are marked by asterisks, and confined flare by squares. Left: non-neutrality factor vs. direct current. Right: SRA vs. its associated magnetic flux. } \label{fig:eruptive-confined}. 	
\end{figure}

To summarize, we confirmed that electric currents flowing through one ribbon are highly correlated with those through the other, hence belonging to the same current system pertinent to the two-ribbon flare. Importantly we found that J-shaped flares are distinct from non-J-shaped flares in a few aspects as follows. First, $|J_z|$, $|J_\mathrm{DC}|$, and $|J_\mathrm{RC}|$ within the SRA of J-shaped flares are significantly smaller than those of non-J-shaped flares, but the SRA of J-shaped flares as well as the associated magnetic flux is significantly larger than that of non-J-shaped flares. In other words, smaller current densities in J-shaped flare ribbons are compensated by larger ribbon area, which explains why J-shaped flares are similar to non-J-shaped ones in terms of electric currents through SRA. Second, the magnitudes of direct, net, and unsigned currents through SRA are all positively correlated with the flare magnitude, but the correlation coefficient for J-shaped flares is generally larger than that for non-J-shaped flares. Third, the majority (27 of 36; 75\%) of J-shaped flares are eruptive, while the majority (30 of 35; 86\%) of non-J-shaped flares are confined; among the 28 hosting active region for the J-shaped flares, 12 are sigmoidal regions \footnote{A special case is AR 11865, which is not yet sigmoidal at the time of Event~\#14 of J-shaped flare, but becomes sigmoidal at the time of Event~\#15.}, yet among the 21 hosting active regions for the non-J-shaped flares, only 3 are sigmoidal. These differences highlight J-shaped flares as a distinct subset of two-ribbon flares, probably the representative of eruptive ones. These differences also imply a different magnetic configuration in J-shaped from non-J-shaped flares, which will be further explored in future work and should be taken into account in developing a three-dimensional model for two-ribbon flares as a whole. 

Furthermore, the flare magnitude is independent of non-neutrality factor $I_\mathrm{nn}$. In other words, it does not matter whether the electric currents flowing through flare ribbons are neutralized ($I_\mathrm{nn}=0$), i.e., balanced separately on each ribbon, or, unneutralized ($I_\mathrm{nn}=1$), i.e., the return current is negligible relative to the direct current through each ribbon. This may imply that both direct and return current are involved in flares. With a weak positive correlation, stronger direct current is associated with $I_\mathrm{nn}$ being closer to unity (left panel of Figure~\ref{fig:eruptive-confined}), but neither direct current nor $I_\mathrm{nn}$ is able to differentiate eruptive from confined flares. We further checked other parameters and any two of their combinations. The only pair that stands out is the SRA and the associated magnetic flux (right panel of Figure~\ref{fig:eruptive-confined}). Most of the flares are eruptive, if SRA exceeds $10^{15}$~m$^2$ and average unsigned magnetic flux through each ribbon exceeds $4\times10^{21}$~Mx. In contrast, if both parameters are below the above-mentioned threshold values, most of the flares are confined.  

Using $|I_\mathrm{DC}/I_\mathrm{RC}|$ to measure the degree of current neutralization, \cite{LiuY2017} found in a sample of four active regions that the two CME-producing regions are strongly nonneutralized ($|I_\mathrm{DC}/I_\mathrm{RC}|\approx2$); the other two regions that did not produce CMEs are almost perfectly neutralized, but can still produce strong flares. They hence speculated that the presence or absence of a double-J-pattern of direct currents around the PIL may indicate whether or not an active region will produce CMEs. More recent studies on larger samples found that CME/flare-productive active regions are more likely to be non-neutralized than quiet regions \citep{Vemareddy2019,Avallone&Sun2020}. In stead of examining the whole active region, our statistical study focuses on electric currents flowing though flare ribbons, in other words, the currents accessible to flares. This study shows that $I_\mathrm{nn}$ is similar to $|I_\mathrm{DC}/I_\mathrm{RC}|$ in its poor correlation with the flare magnitude, consistent with \citet{LiuY2017}, and that J-shaped flares are more likely to be eruptive than non-J-shaped ones. Moreover, we found that combining the ribbon area and its associated magnetic flux has the potential to differentiate eruptive from confined flares. Further study is needed to clarify the connection between electric currents in an active region and its productivity in flares and CMEs.

\acknowledgements  
The authors thank the SDO consortium for the excellent data. This work was supported by NSFC (Grant No. 41761134088, 41774150, and 11925302), CAS Key Research Program (Grant No. KZZD-EW-01-4), the fundamental research funds for the central universities, and the Strategic Priority Program of the Chinese Academy of Sciences (Grant No. XDB41000000)
\newpage



\begin{longrotatetable}
\begin{deluxetable}{cccccccccccccc}	
	\tablecaption{Non-J-shaped flares \label{tab:Non-J-shaped flares}}  	
	\tablehead{
	\colhead{NO.} & \colhead{Flare} & \colhead{Position} & \colhead{NOAA} & \colhead{Category} & \colhead{$\overline{J_z^+}$} & \colhead{$\overline{J_z^-}$} & \colhead{$\widetilde{J_z^+}$} & \colhead{$\widetilde{J_z^-}$} & \colhead{$I_\mathrm{DC}^+$} & \colhead{$I_\mathrm{RC}^+$} & \colhead{$I_\mathrm{DC}^-$} & \colhead{$I_\mathrm{RC}^-$} & \colhead{Noise}\\
	& SXR Peak Time &  & AR &  & mA~m$^{-2}$ & mA~m$^{-2}$ & mA~m$^{-2}$ & mA~m$^{-2}$ & A & A & A & A & mA~m$^{-2}$
	}
    \startdata
 	   		1  & 2011-03-09T14:02 & N9W6   & 11166 & C & 28.98 & 30.03 & 25.84 & 26.76 & 3.10E+11  & -1.59E+11 & -4.13E+11 & 2.18E+11  & 6.67 $\pm$ 5.29  \\
 	   		2  & 2011-04-22T04:57 & S18E43 & 11195 & C & 27.25 & 26.45 & 25.06 & 24.68 & 2.99E+11  & -1.68E+11 & -4.04E+11 & 2.98E+11  & 7.33 $\pm$ 5.48  \\
 	   		3  & 2011-11-06T06:35 & N21E31 & 11339 & C & 40.02 & 35.04 & 30.42 & 29.68 & -2.23E+12 & 1.61E+12  & 1.83E+12  & -8.33E+11 & 7.87 $\pm$ 6.00  \\
 	   		4  & 2011-12-26T02:27 & S21W33 & 11387 & E & 36.96 & 35.01 & 32.33 & 29.37 & -1.33E+12 & 7.42E+11  & 1.25E+12  & -4.77E+11 & 10.12 $\pm$ 7.99 \\
 	   		5  & 2011-12-31T16:26 & S26E42 & 11389 & C & 35.29 & 40.32 & 29.53 & 30.80 & -8.65E+11 & 6.69E+11  & 1.80E+12  & -5.04E+11 & 8.44 $\pm$ 6.66  \\
 	   		6  & 2012-03-06T00:28 & N16E41 & 11429* & C & 38.32 & 45.62 & 32.45 & 36.34 & -1.70E+12 & 1.55E+12  & 4.61E+12  & -1.47E+12 & 8.99 $\pm$ 6.97  \\
 	   		7  & 2012-03-06T01:44 & N17E41 & 11429* & C & 41.11 & 44.64 & 34.29 & 34.23 & -2.96E+12 & 1.69E+12  & 5.26E+12  & -1.77E+12 & 9.15 $\pm$ 7.05  \\
 	   		8  & 2012-03-06T22:53 & N17E35 & 11429* & C & 40.85 & 50.77 & 34.16 & 39.45 & -3.68E+12 & 1.31E+12  & 6.65E+12  & -1.58E+12 & 8.81 $\pm$ 6.74  \\
 	   		9  & 2012-03-06T12:41 & N18E36 & 11429* & C & 34.96 & 46.73 & 30.29 & 37.24 & -3.16E+12 & 1.51E+12  & 5.78E+12  & -1.39E+12 & 10.54 $\pm$ 7.88 \\
 	   		10 & 2012-06-06T20:06 & S19W5  & 11494 & E & 30.80 & 33.40 & 27.45 & 28.30 & -6.31E+11 & 5.56E+11  & 9.60E+11  & -4.01E+11 & 9.05 $\pm$ 6.87  \\
 	   		11 & 2012-06-13T13:17 & S16E18 & 11504* & E & 34.83 & 37.93 & 28.59 & 31.25 & 1.28E+12 & -9.24E+11 & -3.54E+12 & 1.72E+12 & 7.01 $\pm$ 5.54  \\
 	   		12 & 2012-07-14T04:58 & S22W30 & 11521 & C & 33.40 & 31.86 & 28.00 & 29.97 & 7.47E+11  & -3.67E+11 & -1.07E+12 & 2.50E+11  & 7.56 $\pm$ 5.98  \\
 	   		13 & 2012-11-27T21:26 & S14W41 & 11620 & C & 45.00 & 35.05 & 33.97 & 29.88 & 3.03E+12  & -1.14E+12 & -1.34E+12 & 8.72E+11  & 9.04 $\pm$ 7.07  \\
 	   		14 & 2013-08-12T10:41 & S17E19 & 11817* & E & 36.53 & 50.98 & 31.57 & 37.60 & -1.61E+12 & 3.51E+11  & 3.05E+12  & -6.00E+11 & 9.71 $\pm$ 7.59  \\
 	   		15 & 2013-10-22T00:22 & N6E17  & 11875 & C & 37.33 & 35.00 & 31.24 & 30.13 & 1.26E+12  & -1.24E+12 & -1.03E+12 & 7.26E+11  & 9.67 $\pm$ 7.61  \\
 	   		16 & 2013-10-24T00:08 & N8W11  & 11875 & C & 39.12 & 40.47 & 32.38 & 31.89 & 2.58E+12  & -1.73E+12 & -3.00E+12 & 2.05E+12  & 8.56 $\pm$ 6.41  \\
 	   		17 & 2013-10-24T10:09 & N6W14  & 11875 & C & 36.56 & 40.75 & 31.68 & 31.37 & 1.93E+12  & -1.27E+12 & -2.34E+12 & 1.45E+12  & 6.13 $\pm$ 4.83  \\
 	   		18 & 2013-11-16T04:53 & S19W29 & 11900 & C & 39.82 & 46.78 & 32.97 & 34.14 & -1.76E+12 & 1.78E+11  & 1.87E+12  & -2.20E+11 & 7.80 $\pm$ 6.28  \\
 	   		19 & 2014-02-01T01:25 & S11E26 & 11967 & C & 33.61 & 54.77 & 28.59 & 40.56 & -1.83E+12 & 7.78E+11  & 3.17E+12  & -1.63E+12 & 7.07 $\pm$ 5.66  \\
 	   		20 & 2014-02-02T22:04 & S13E5  & 11967 & C & 41.83 & 42.75 & 35.92 & 34.25 & -3.87E+12 & 1.69E+12  & 3.57E+12  & -1.53E+12 & 7.64 $\pm$ 6.62  \\
 	   		21 & 2014-02-04T01:23 & S13W14 & 11967 & C & 48.31 & 61.78 & 38.07 & 44.68 & -4.42E+12 & 8.43E+11  & 4.73E+12  & -5.35E+11 & 9.49 $\pm$ 7.40  \\
 	   		22 & 2014-02-04T09:49 & S13W12 & 11967 & C & 50.76 & 54.57 & 40.46 & 40.12 & -2.92E+12 & 8.55E+11  & 3.40E+12  & -5.79E+11 & 8.70 $\pm$ 6.70  \\
 	   		23 & 2014-02-11T16:51 & S13E12 & 11974 & E & 36.11 & 38.15 & 31.12 & 30.31 & 2.31E+12  & -1.38E+12 & -2.30E+12 & 8.90E+11  & 7.65 $\pm$ 6.29  \\
 	   		24 & 2014-02-13T08:12 & S12W13 & 11974 & C & 46.40 & 41.96 & 34.43 & 33.43 & 3.31E+12  & -2.71E+12 & -2.66E+12 & 2.56E+12  & 9.04 $\pm$ 7.14  \\
 	   		25 & 2014-10-20T09:11 & S13E43 & 12192 & C & 39.38 & 44.86 & 32.87 & 33.12 & -2.60E+12 & 1.88E+12  & 5.16E+12  & -3.20E+12 & 8.33 $\pm$ 6.40  \\
 	   		26 & 2014-10-20T20:04 & S13E43 & 12192 & C & 31.65 & 36.85 & 28.04 & 32.86 & -7.64E+11 & 6.68E+11  & -6.55E+11 & 5.36E+11  & 8.32 $\pm$ 6.49  \\
 	   		27 & 2014-11-07T10:22 & N15E43 & 12205 & C & 46.59 & 37.01 & 36.24 & 30.78 & 2.50E+12  & -7.27E+11 & -1.50E+12 & 7.26E+11  & 11.14 $\pm$ 8.36 \\
 	   		28 & 2014-12-01T06:41 & S21E17 & 12222 & C & 28.58 & 29.31 & 27.19 & 26.59 & -5.97E+11 & 2.68E+11  & 6.73E+11  & -2.98E+11 & 10.23 $\pm$ 7.77 \\
 	   		29 & 2014-12-04T18:25 & S20W32 & 12222 & C & 36.07 & 33.92 & 29.82 & 28.11 & 3.12E+12  & -3.08E+12 & 2.19E+12  & -1.74E+12 & 9.68 $\pm$ 7.57  \\
 	   		30 & 2014-12-04T19:41 & S20W32 & 12222 & C & 32.81 & 25.76 & 30.01 & 24.70 & -4.89E+11 & 1.54E+11  & -1.79E+11 & 6.00E+10  & 8.46 $\pm$ 6.69  \\
 	   		31 & 2014-12-05T12:25 & S19W37 & 12222 & C & 31.72 & 29.36 & 27.67 & 26.11 & 1.21E+12  & -8.91E+11 & 8.47E+11  & -4.81E+11 & 9.03 $\pm$ 6.99  \\
 	   		32 & 2014-12-19T09:44 & S19W27 & 12242 & C & 29.66 & 28.30 & 27.58 & 25.75 & 5.17E+11  & -3.67E+11 & 4.75E+11  & -2.80E+11 & 7.85 $\pm$ 6.18  \\
 	   		33 & 2014-12-21T07:32 & S19W44 & 12242 & C & 38.49 & 34.61 & 32.19 & 29.55 & 2.21E+12  & -1.10E+12 & -1.29E+12 & 1.17E+12  & 8.10 $\pm$ 6.38  \\
 	   		34 & 2015-01-26T16:53 & S10E25 & 12268 & C & 37.74 & 39.66 & 30.47 & 32.43 & -1.00E+12 & 5.68E+11  & 1.86E+12  & -1.09E+12 & 9.62 $\pm$ 7.40  \\
 	   		35 & 2015-06-20T06:48 & N13E27 & 12371 & C & 42.39 & 44.21 & 30.17 & 33.78 & -1.28E+12 & 2.86E+11  & 8.87E+11  & -6.07E+11 & 10.73 $\pm$ 8.13   	
        \enddata
        \tablecomments{	`C' and `E' indicate confined and eruptive flares, respectively, in the `Category' column. Sigmoidal active regions are marked by asterisks.		
        }
\end{deluxetable}
\end{longrotatetable}

\begin{longrotatetable}
	\begin{deluxetable}{cccccccccccccc}	
		
	\tablecaption{J-shaped flares \label{tab:J-shaped flares}}  	
	\tablehead{
\colhead{No.} & \colhead{Flare} & \colhead{Position} & \colhead{NOAA} & \colhead{Category} & \colhead{$\overline{J_z^+}$} & \colhead{$\overline{J_z^-}$} & \colhead{$\widetilde{J_z^+}$} & \colhead{$\widetilde{J_z^-}$} & \colhead{$I_\mathrm{DC}^+$} & \colhead{$I_\mathrm{RC}^+$} & \colhead{$I_\mathrm{DC}^-$} & \colhead{$I_\mathrm{RC}^-$} & \colhead{Noise}\\
& SXR Peak Time &  & AR &  & mA~m$^{-2}$ & mA~m$^{-2}$ & mA~m$^{-2}$ & mA~m$^{-2}$ & A & A & A & A & mA~m$^{-2}$	
	}
       \startdata
      		1  & 2010-08-07T18:24 & N11E34 & 11093* & E & 26.16 & 25.77 & 24.83 & 24.92 & -6.74E+11 & 5.81E+11  & 6.05E+11  & -4.20E+11 & 8.46 $\pm$ 6.46  \\
       		2  & 2011-07-27T16:07 & N20E37 & 11260 & C & 44.39 & 33.95 & 32.26 & 29.15 & 3.07E+12  & -2.53E+12 & 2.39E+12  & -1.63E+12 & 11.73 $\pm$ 8.64 \\
       		3  & 2011-08-02T06:19 & N16W8  & 11261 & E & 45.68 & 42.81 & 32.82 & 33.81 & 4.28E+12  & -8.23E+11 & -4.54E+12 & 1.94E+12  & 6.62 $\pm$ 5.50  \\
       		4  & 2011-08-03T13:48 & N16W30 & 11261 & E & 38.88 & 35.78 & 30.92 & 29.52 & 3.48E+12  & -9.35E+11 & -2.70E+12 & 1.23E+12  & 5.98 $\pm$ 4.62  \\
       		5  & 2011-11-09T13:35 & N25E34 & 11342 & E & 26.93 & 26.66 & 25.72 & 25.49 & -5.50E+12 & 5.44E+12  & -3.48E+12 & 3.14E+12  & 10.38 $\pm$ 7.74 \\
       		6  & 2012-03-10T17:44 & N18W18 & 11429 & E & 30.79 & 30.18 & 27.17 & 27.30 & -3.06E+12 & 1.63E+12  & 3.87E+12  & -1.86E+12 & 8.61 $\pm$ 6.74  \\
       		7  & 2012-03-14T15:21 & N14E5  & 11432 & E & 28.82 & 30.13 & 26.83 & 26.87 & 8.59E+11  & -4.22E+11 & -6.12E+11 & 4.76E+11  & 7.28 $\pm$ 5.78  \\
       		8  & 2012-04-27T08:24 & N11W30 & 11466* & C & 27.90 & 29.52 & 25.89 & 26.53 & -5.11E+11 & 4.04E+11  & 8.45E+11  & -5.17E+11 & 9.22 $\pm$ 7.24  \\
       		9  & 2012-07-10T05:14 & S17E33 & 11520 & C & 34.37 & 49.04 & 29.38 & 41.64 & 9.58E+11  & -3.38E+11 & -2.33E+12 & 4.16E+11  & 7.23 $\pm$ 5.92  \\
       		10 & 2012-07-10T06:27 & S17E30 & 11520 & C & 34.11 & 39.05 & 30.13 & 34.74 & 1.00E+12  & -4.01E+11 & -1.57E+12 & 1.46E+11  & 8.77 $\pm$ 6.73  \\
       		11 & 2013-04-11T07:16 & N9E12  & 11719* & E & 27.15 & 27.10 & 25.99 & 25.87 & -2.60E+12 & 2.40E+12  & 1.02E+12  & -9.50E+11 & 10.35 $\pm$ 7.73 \\
       		12 & 2013-04-12T20:38 & N22W42 & 11718 & C & 30.32 & 34.79 & 26.69 & 30.13 & 4.71E+11  & -3.74E+11 & -1.33E+12 & 1.11E+12  & 9.05 $\pm$ 7.43  \\
       		13 & 2013-08-17T19:33 & S7W29  & 11818* & E & 39.30 & 29.16 & 31.43 & 26.34 & 1.66E+12  & -6.56E+11 & -1.61E+12 & 1.38E+12  & 10.27 $\pm$ 8.23 \\
       		14 & 2013-10-13T00:43 & S22E17 & 11865 & E & 38.67 & 34.96 & 29.42 & 31.25 & 2.11E+12  & -3.79E+11 & -1.78E+12 & 5.49E+11  & 10.03 $\pm$ 7.69 \\
       		15 & 2013-10-15T08:38 & S22W13 & 11865* & E & 62.40 & 42.13 & 36.87 & 33.01 & 2.58E+12  & -2.69E+11 & -1.91E+12 & 2.92E+11  & 8.32 $\pm$ 6.40  \\
       		16 & 2014-01-07T18:32 & S14W7  & 11944 & E & 29.16 & 27.48 & 26.07 & 25.68 & 1.36E+12  & -1.19E+12 & -2.73E+12 & 2.51E+12  & 8.31 $\pm$ 6.53  \\
       		17 & 2014-01-31T15:42 & N9E36  & 11968 & E & 27.95 & 28.00 & 26.60 & 26.49 & 2.57E+12  & -2.39E+12 & -2.14E+12 & 1.90E+12  & 12.56 $\pm$ 9.45 \\
       		18 & 2014-02-01T07:23 & S11E23 & 11967 & C & 35.37 & 47.78 & 29.21 & 36.78 & -1.50E+12 & 6.58E+11  & 3.67E+12  & -1.43E+12 & 8.18 $\pm$ 7.08  \\
       		19 & 2014-03-20T03:56 & S14E35 & 12010 & E & 27.96 & 31.87 & 26.38 & 27.21 & -4.49E+11 & 3.12E+11  & -5.50E+11 & 3.94E+11  & 7.82 $\pm$ 6.27  \\
       		20 & 2014-06-16T00:01 & S22E7  & 12087 & E & 27.65 & 28.35 & 25.93 & 25.38 & -4.70E+11 & 4.35E+11  & -5.11E+11 & 4.37E+11  & 8.48 $\pm$ 6.69  \\
       		21 & 2014-08-01T18:13 & S9E12  & 12127 & E & 29.22 & 28.47 & 26.63 & 25.53 & -1.02E+12 & 6.58E+11  & 7.45E+11  & -4.24E+11 & 6.95 $\pm$ 5.37  \\
       		22 & 2014-08-25T15:11 & N5W36  & 12146 & E & 31.37 & 45.25 & 28.01 & 35.50 & -1.46E+12 & 4.83E+11  & 2.07E+12  & -5.76E+11 & 6.75 $\pm$ 5.35  \\
       		23 & 2014-08-25T20:21 & N9W38  & 12146 & E & 30.86 & 39.91 & 28.80 & 30.56 & -1.81E+12 & 5.60E+11  & 2.27E+12  & -3.58E+11 & 7.52 $\pm$ 6.02  \\
       		24 & 2014-09-09T00:29 & N12E29 & 12158* & E & 34.83 & 34.11 & 30.14 & 29.05 & -2.66E+12 & 1.50E+12  & 2.50E+12  & -1.29E+12 & 8.63 $\pm$ 6.59  \\
       		25 & 2014-09-28T02:58 & S13W23 & 12173* & E & 33.31 & 29.67 & 28.33 & 26.77 & -1.58E+12 & 1.15E+12  & 9.70E+11  & -4.65E+11 & 8.12 $\pm$ 6.23  \\
       		26 & 2014-10-24T21:41 & S16W21 & 12192* & C & 33.19 & 30.54 & 27.29 & 26.88 & -6.07E+12 & 4.73E+12  & 4.52E+12  & -3.36E+12 & 8.21 $\pm$ 6.35  \\
       		27 & 2014-11-07T17:26 & N15E33 & 12205 & E & 46.31 & 35.80 & 30.97 & 29.95 & 7.78E+12  & -1.60E+12 & -5.02E+12 & 1.92E+12  & 8.19 $\pm$ 6.51  \\
       		28 & 2014-12-17T19:01 & S10E24 & 12241 & C & 37.50 & 37.16 & 30.06 & 31.31 & 1.26E+12  & -2.71E+11 & -1.66E+12 & 8.66E+11  & 9.81 $\pm$ 7.56  \\
       		29 & 2014-12-18T21:58 & S15E8  & 12241 & E & 37.02 & 36.65 & 31.02 & 32.10 & 3.35E+12  & -5.08E+11 & -4.39E+12 & 1.04E+12  & 9.69 $\pm$ 7.57  \\
       		30 & 2014-12-21T12;17 & S11W21 & 12241 & E & 32.45 & 27.36 & 26.76 & 25.72 & 6.35E+11  & -2.60E+11 & -5.53E+11 & 2.85E+11  & 8.16 $\pm$ 6.36  \\
       		31 & 2015-03-12T12:14 & S16E6  & 12297 & C & 35.59 & 41.53 & 29.54 & 28.62 & 9.92E+11  & -2.77E+11 & -1.07E+12 & 4.60E+11  & 7.54 $\pm$ 5.98  \\
       		32 & 2015-06-21T02:36 & N13E14 & 12371* & E & 37.24 & 34.42 & 29.87 & 28.73 & -1.70E+12 & 4.72E+11  & 1.36E+12  & -9.95E+11 & 9.52 $\pm$ 7.25  \\
       		33 & 2015-06-21T01:42 & N12E13 & 12371* & E & 38.77 & 33.72 & 29.92 & 30.27 & -2.37E+12 & 7.99E+11  & 2.19E+12  & -1.08E+12 & 8.21 $\pm$ 6.31  \\
       		34 & 2015-06-22T18:23 & N12W8  & 12371* & E & 30.88 & 30.92 & 28.17 & 27.18 & -3.01E+12 & 6.49E+11  & 2.97E+12  & -1.22E+12 & 7.62 $\pm$ 6.02  \\
       		35 & 2015-11-04T13:52 & N6E3   & 12443 & E & 29.62 & 28.18 & 26.75 & 26.16 & 1.72E+12  & -1.22E+12 & -1.44E+12 & 1.43E+12  & 6.83 $\pm$ 5.56  \\
       		36 & 2015-11-09T13:12 & S11E41 & 12449* & E & 27.87 & 28.42 & 26.24 & 26.44 & -1.55E+12 & 1.48E+12  & 2.74E+12  & -1.94E+12 & 9.81 $\pm$ 7.42 
       	            \enddata
	
\tablecomments{	`C' and `E' indicate confined and eruptive flares, respectively, in the `Category' column. Sigmoidal active regions are marked by asterisks.		
}
	
\end{deluxetable}
\end{longrotatetable}

\end{document}